\documentstyle [preprint,aps,epsf]{revtex}
\begin{document}
\title{Single-particle properties of a model for coexisting
charge and spin quasi-critical fluctuations coupled to
electrons}
\author{S. Caprara, M. Sulpizi, A. Bianconi, C. Di Castro, and M. Grilli} 
\address {Dipartimento di Fisica - Universit\`a di Roma ``La Sapienza'' \\
and Istituto Nazionale di Fisica della Materia, Unit\`a di Roma 1 \\
P.le A. Moro, 2 - 00185 Roma - Italy}
\maketitle

\begin{abstract}
We study the single-particle spectral properties of a model for coexisting AFM 
and ICDW critical fluctuations coupled to electrons, which naturally arises in 
the context of the stripe-quantum-critical-point scenario for high-$T_c$ 
superconducting materials. Within a perturbative approach, we show that the 
on-shell inverse scattering time deviates from the normal Fermi-liquid 
behavior near the points of the Fermi surface connected by the characteristic 
wave-vectors of the critical fluctuations (hot spots). The anomalous behavior 
is stronger when the hot spots are located near singular points of the
electronic spectrum.

The violations to the normal Fermi-liquid behavior are associated with the 
transfer of spectral weight from the quasi-particle peak to incoherent 
shadow peaks, which produces an enhancement of incoherent spectral weight
near the Fermi level.

We use our results to discuss recent ARPES experiments on Bi2212 near optimal
doping. 
\end{abstract}
\vskip 1truecm

\section{Introduction}
The nature of the single-particle excitations in a metal close to a Quantum 
Critical Point (QCP) has become a subject of interest after the proposal that 
the anomalous properties of the metallic phase of high-$T_c$ superconductors 
may be related to the presence of a singular effective scattering amplitude, 
such as the one arising near a QCP of some kind of instability, in the Random 
Phase Approximation (RPA). Two realization of this scenario have been proposed, 
the first associated with an antiferromagnetic (AFM) instability 
\cite{sach,NAFL,pines}, the second with an incommensurate charge-density-wave 
(ICDW) instability \cite{cdg1}.

The presence of an AFM phase at zero and very low doping and of strong AFM 
fluctuations, revealed in neutron-scattering experiments at higher doping 
\cite{neutron}, has suggested that the relevant physics of high-$T_c$
superconducting materials is dominated by the AFM QCP. However in such scenario 
the QCP itself is quite far from the point of optimal doping where both the 
highest $T_c$ and the maximum violations of the Fermi-liquid (FL) behavior in 
the metallic phase are observed. To explain this evident inconsistency strong 
vertex corrections have to be advocated in order to suppress the effect of AFM 
fluctuations \cite{chubukov}, which would otherwise be strongest at lower 
doping, close to the QCP. Moreover the peculiar role of the optimal-doping
point in all classes of high-$T_c$ materials is left unexplained.

To avoid this problem, and considering also that previous theoretical results
indicate the presence of a charge instability, the existence of a new QCP
was proposed, which controls the physics of the cuprates near optimal doping.
Indeed, models for strongly correlated electrons with short-range interactions 
are commonly characterized by an instability with respect to phase separation 
\cite{ps}, which is turned into an instability with respect to ICDW when 
long-range Coulomb forces are taken into account \cite{cdg1,becca}.

The most direct evidence for a new QCP at (or near to) optimal doping is 
provided by resistivity measurements \cite{boebinger}, which reveal an 
insulator-to-metal transition when the superconducting state is suppressed by 
a strong pulsed magnetic field. Neutron scattering experiments evidenced 
charge-driven order in related compounds \cite{tranq}. Coexistence of ICDW and 
electron gas has been detected in joint EXAFS \cite{exafs,xxx} 
and X-ray diffraction \cite{ics} experiments (see also Ref. \cite{ab}).

The new QCP is not incompatible with an AFM QCP. The two coexist and determine
the physics of high-$T_c$ superconducting materials in different regions of 
the phase diagram. Near optimal doping it is very likely that charge degrees 
of freedom play a major role. Moreover the dynamical charge segregation into
hole-rich and hole-poor regions, associated with critical charge fluctuations, 
would enhance AFM fluctuations far away from the AFM QCP, due to the natural 
tendency of hole-poor regions towards antiferromagnetism. In this way a 
consistent framework is achieved to understand the role of spin degrees of 
freedom in optimally doped and overdoped materials. AFM fluctuations enslaved 
by the onset of charge dynamical fluctuations were evidenced in related 
compounds \cite{tranq}.

Different approaches have been used so far to capture the relevant features of 
the single-particle excitation spectra in a metal close to a QCP. In the 
mean-field theory of spin-density-wave (SDW) antiferromagnetism the 
quasi-particle spectrum may be obtained by coupling the electrons to spin 
fluctuations characterized by a factorized susceptibility 
$\chi=i\pi S(S+1)\delta(\omega)\delta(q-Q)$, where $Q=(\pi/a,...,\pi/a)$ is 
the wave-vector of the AFM structure and $S$ is the spin quantum number of 
the fluctuating spin. Within this approach the gap in the 
quasi-particle spectrum is proportional to the effective coupling $g$. Kampf 
and Schrieffer proposed that the resulting self-energy
$\Sigma (k,\varepsilon)=g^2[(\varepsilon-\xi_{k-Q})^{-1}-i\pi
{\rm sgn}(\varepsilon) \delta(\varepsilon-\xi_{k-Q})]$, could be slightly
modified, near the AFM QCP, due to the interaction between free
electrons and AFM fluctuations \cite{ks,tagliacozzo}, 
which leads to a broadening 
$|{\rm Im}\Sigma| \simeq g^2\Gamma/[(\varepsilon-\xi_{k-Q})^2+\Gamma^2]$
so that ${\rm Re}\Sigma\simeq g^2(\varepsilon-\xi_{k-Q})/[
(\varepsilon-\xi_{k-Q})^2+\Gamma^2]$ has a quasi-polar structure \cite{notaks}. 
Thus the inverse scattering time at the Fermi energy is finite 
$1/\tau\simeq 2\Gamma/[\xi_{k-Q}^2+\Gamma^2]$ and new (quasi-)poles appear in 
the electron Green function at energies $\varepsilon\sim\xi_{k-Q}$, which are 
usually called shadow bands.

Within the framework of the nearly-AFM FL (NAFL) \cite{NAFL}, Chubukov 
{\sl et al.} \cite{chubukov}, investigated the single-particle properties of 
a model for  electrons coupled to spin fluctuations characterized by a 
phenomenological susceptibility of the form proposed by Millis, Monien and 
Pines \cite{millis}. An undamped susceptibility $\chi(q,\omega)$ was used to 
describe mainly the underdoped region. Within perturbation theory a topological 
transition of the Fermi surface (FS) was found at a finite value of the 
electron-spin fluctuations coupling constant \cite{chubukov}, leading to the 
formation of hole pockets located around the points $(\pm \pi/2a,\pm \pi/2a)$
of the Brillouin zone. Such a transition is associated with the appearance of
new poles in the electron Green function which are the result of the evolution
of the electron self-energy towards a mean-field-like behavior similar to that 
suggested in Ref. \cite{ks,tagliacozzo}. 
Thus, in this scenario, the precursors of the AFM 
phase show up in the anomalous metallic phase, at higher doping, with a change 
in the nature of the quasi-particles, signalling the incipient 
antiferromagnetism.

However the experimental results provide an increasing evidence for the 
existence of stripe phases with modulations incommensurate with the lattice
\cite{stripes98}.
This makes at least questionable a description in terms of AFM fluctuations 
only. Moreover, the topology of the FS of optimally doped Bi2212 was recently 
determined with great accuracy \cite{bianconi}, revealing new relevant 
features, and namely an asymmetric suppression of the spectral weight near the 
Fermi level around the $M$ points of the Brillouin zone, at points of the FS
connected by an incommensurate wave-vector $Q_c=(0.4\pi/a,-0.4\pi/a)$, in
addition to a symmetric suppression at the same points, associated with the
wavevector $Q_s=(\pi/a,\pi/a)$.

This may be interpreted within the ICDW QCP scenario \cite{cdg1}, as a 
signature of quasi-critical charge fluctuations close to a stripe phase. In 
the following we want to check the consistency of this interpretation, and
suggest that ARPES experiments can be used to extract the properties of the 
charge-fluctuation spectrum, from the related features appearing in the 
single-particle excitation spectra.

Since we expect that the onset of a stripe phase reintroduces AFM fluctuations,
we consider as a starting point an effective model for coexisting charge
and spin fluctuations coupled to conduction electrons with bare band dispersion 
$\xi_k$. The properties of such charge and spin fluctuations are generically
described within a RPA around the QCP at optimal doping, which can be considered
as the point governing the onset of the stripe phase. The explicit form of the
parameters appearing in the resulting charge and spin susceptibilities, which
depend on the underlying microscopic model, will be deduced below by means
of a phenomenological analysis, mainly based on ARPES experiments. In 
particular, the modulation of the density waves associated with charge and
spin fluctuations is characterized by the wave-vectors $Q_c$ and $Q_s$ 
respectively.

Within a perturbative approach we show that such a model exhibits 
a violation of the normal FL behavior associated with the transfer of spectral 
weight from the quasi-particle peak at an energy $\varepsilon\simeq \xi_k$ to 
dispersing incoherent bands located at $\varepsilon\simeq\xi_{k-Q_{c,s}}$, 
without the appearance of new poles in the electron Green function. The 
resulting bands have thus the aspect, but not the coherent nature of the 
shadow bands found within the approaches of Refs. \cite{chubukov,ks}. 
Such bands could be related to corresponding broad features seen in ARPES
experiments, such as the peak at an energy $\varepsilon\simeq -200$ meV,
observed near the $M$ points of the Brillouin zone in Bi2212 \cite{marshall}.

We point out that the shadow features of incoherent nature were also found by 
Bennemann {\sl et al.} within the FLEX approximation for the Hubbard model 
\cite{bennemann}. The interpretation of the shadow bands as spectral-weight 
anomalies \cite{ss}
was also suggested in ref. \cite{larosa}. More recently 
similar results were found within the NAFL scenario \cite{schpi}, by performing
a full re-summation of the entire perturbation series in the high-temperature 
(static) limit for the spin-fluctuation spectrum. We point out that these 
results agree with our simpler perturbative results, in the case of electrons 
coupled to spin fluctuations only. Moreover, by considering also the effect of 
charge fluctuations, we reproduce new features, which have a counterpart in
recent ARPES experiments, and call for further investigation.

The scheme of the paper is the following. In Sec. II we introduce the model for 
tight-binding electrons coupled to charge and spin fluctuations. In Sec. III 
we discuss the violation of the FL behavior for the on-shell inverse scattering 
time. In Sec. IV we analyze the single-particle spectral properties of our 
model. Sec. V is devoted to the comparison between our results and ARPES 
experiments on Bi2212 near optimal doping. 
Concluding remarks are found in Sec. VI.

\section{The model}

We consider a model for tight-binding electrons coupled to charge and spin 
fluctuations, described by the Hamiltonian
\begin{equation}
{\cal H}=\sum_{k,\sigma}\left(\xi_k-\delta\mu\right) 
c^{+}_{k,\sigma}c_{k,\sigma}+
\sum_i g_i\sum_{kq,\alpha\beta}c^{+}_{k+q,\alpha}c_{k,\beta}
\tau^i_{\alpha\beta}S^i_{-q},
\label{hamilt}
\end{equation}
where $\xi_k=-2t[\cos(k_{x}a)+\cos(k_{y}a)]-4t'\cos(k_{x}a)\cos(k_{y}a)-\mu$
is the dispersion law for free electrons on a square bi-dimensional 
lattice for the CuO$_2$ planes with hopping terms up to next-to-nearest 
neighbors included, $a$ is the lattice spacing, and $\mu$ is the bare chemical 
potential. In the following, for the sake of definiteness, and aiming in the end
for a comparison with ARPES experiments, we assume that $-1<2t'/t<0$, to fix 
the main properties of the bare band structure and the shape of the FS, as
appropriate to the Bi2212 samples. The correction $\delta\mu\sim O(g^2)$ to the
chemical potential is treated perturbatively and is fixed to ensure that the 
number of particles is the same as that in the free-electron system with bare 
chemical potential $\mu$.

The second term in (\ref{hamilt}) describes the interaction of electrons with 
charge ($i=0$) and spin ($i=1,2,3$) fluctuating fields $S^i_{-q}$, which is 
characterized by the coupling constants $g_i$. The spin structure of the 
generalized electron density coupled to the fluctuating field $S_{-q}^i$ is 
given by the corresponding Pauli matrix $\tau^i$. We assume that the 
phenomenological properties of the fluctuating fields $S_{-q}^i$ are completely 
described by correlation functions of the form
\begin{equation}
\chi_{ij}(q,\omega)=\frac{\delta_{ij} A_j}{\Omega_j(q)-i\omega},
\label{chi}
\end{equation}
where, for each $j$, $A_j$ is some constant, $\Omega_j(q)=
M_j+\alpha_j T+{\nu}_j \gamma_{q-Q_j}$ with $\gamma_q=
2-\cos(q_x a)-\cos(q_y a)$, $Q_j$ is the characteristic wave-vector
of the critical fluctuations, ${\nu}_j$ is the inverse of the characteristic 
time scale of the fluctuating mode and the distance from criticality is 
measured by the ``mass term'' $M_j+\alpha_j T$. In particular $M_j$ is the
distance from the QCP at zero temperature, which we assume to depend on the 
hole doping per unit cell $x$ and to vanish at a critical value $x_c$, and
$T$ is the temperature in energetic units, so that $\alpha_j$ is a 
dimensionless constant. We point out that the cos-like form of the dispersion 
in $\gamma(q)$ is adopted to reproduce a $q^2$ behavior at small momenta, 
while preserving the lattice periodicity \cite{cdg1}.

We further assume complete spin isotropy, so that all the above parameters are 
the same for $j=1,2,3$. Thus in the following we turn to the notation
$j=c,s$ for all the constants related to charge and spin fluctuating
fields respectively. We also introduce the dimensionless coupling constants 
$\lambda_c=g_c^2 A_c/t{\nu}_c$ and $\lambda_s=g_s^2 A_s/t{\nu}_s$.

In the case of AFM fluctuations the susceptibility (\ref{chi}) corresponds
to the phenomenological expression obtained by Millis, Monien and Pines
\cite{millis} in the limit of strong damping; in the case of ICDW a similar 
expression was found close to the instability in the Hubbard-Holstein model 
with long-range Coulomb forces, within a slave-boson approach \cite{cdg1,becca}.
In the present phenomenological approach the fact that charge instability
leads to an enslaved spin modulation at the stripe QCP, is modelled by
the requirement that the vanishing of the mass $M_s$ is guided by the
vanishing of the mass $M_c$. The determination of explicit dependence of 
$M_s$ on $M_c$ requires the introduction of a specific microscopic model,
which is beyond the scope of this paper. We point out that, in the static limit 
$\omega\to 0$, and for momenta $q\ll 1/a$, the susceptibilities (\ref{chi}) 
have the form of the Ornstein-Zernicke critical correlation function.

Close to criticality ($T,M_j\ll t$) a perturbative analysis is in
principle not justified.
Nonetheless, to capture the essential features of the single-particle spectra 
when electrons are coupled through a singular effective interaction, we
perform our calculations within perturbation theory. The agreement with the
full calculations \cite{schpi} in the static limit, as well as the claimed
smallness of vertex corrections \cite{chubukov}, provide a support to the
substantial validity of our approach. Since a full microscopic derivation of the
parameters appearing in our model would not be appropriate in this context,
we rather follow a phenomenological analysis, based on the interpretation of
experimental results.

The lowest order in perturbation theory gives an electron self-energy
\begin{equation} 
\Sigma(k,\varepsilon)=
\Sigma_c (k,\varepsilon)+3\Sigma_s (k,\varepsilon)-\delta\mu,
\label{selfene}
\end{equation}
where a factor of 3 in front of $\Sigma_s$ comes out from the sum over the 
three spin components $j=1,2,3$, which are equivalent in the paramagnetic 
phase. We discuss the single-particle spectral properties by means of the
spectral density 
\begin{equation}
A(k,\varepsilon)=\frac{1}{\pi}\frac{|{\rm Im}\Sigma(k,\varepsilon)|}
{[\varepsilon-\xi_{k}-{\rm Re}\Sigma(k,\varepsilon)]^2+
[{\rm Im}\Sigma(k,\varepsilon)]^2}.
\label{ak}
\end{equation}
The correction $\delta\mu$ to the chemical potential is fixed, for a given set 
of coupling constants $\{\lambda_j\}$, by the condition
\begin{equation}
2 \int_{-\pi/a}^{\pi/a}\int_{-\pi/a}^{\pi/a}\frac{d^2k}{(2\pi/a)^2}
\int_{-\infty}^{+\infty} 
A(k,\varepsilon)f(\varepsilon)d\varepsilon =1-x,
\label{pchim}
\end{equation}
where $x$ is the hole doping per unit cell with respect to half filling,
and a factor of 2 in the l.h.s. accounts for spin degeneracy. The self-energies
appearing in (\ref{selfene}) are calculated at a fixed value of the bare
chemical potential $\mu$, which corresponds to the same hole doping $x$. The
explicit expression for the imaginary part of the self-energies 
appearing in (\ref{selfene}) is 
\begin{equation}
{\rm Im}\Sigma_j(k,\varepsilon)=\lambda_j {\nu}_j t
{\rm sgn}(\varepsilon)
\int_{-\pi/a}^{\pi/a}\int_{-\pi/a}^{\pi/a}\frac{d^2k'}{(2\pi/a)^2}
\frac{[\varepsilon-\xi_{k'}][f(\xi_{k'})+b(\xi_{k'}-\varepsilon)]}
{[\varepsilon-\xi_{k'}]^2+\Omega_j^2(k-k')},
\label{imsigma}
\end{equation}
where $f(\varepsilon)=[e^{\varepsilon/T}+1]^{-1}$ is the Fermi 
function, and $b(\varepsilon)=[e^{\varepsilon/T}-1]^{-1}$ is the Bose function.
Within our approach the momentum cut-off $q_{max}=\pi/a$ is provided by the
underlying lattice. Since, however, we assumed the specific form (\ref{chi})
for the susceptibilities, which is expected to hold for momenta $q\simeq Q_j$,
we introduce in (\ref{imsigma}) 
a smooth cut-off function ${\cal C}_{\Delta}(k-k'-Q_j)=
\exp(-\gamma_{k-k'-Q_j}/\Delta^2)$, whenever a quantitative improvement is 
needed to compare with experimental results. We point out that the
presence of a finite cut-off $\Delta$ does not change the qualitative behavior
of the quantities under discussion.

At $T=0$ the integral over $k'$ in (\ref{imsigma}) is restricted to the region
$0\le\xi_{k'}\le\min(\varepsilon,\xi_{max})$ for $\varepsilon>0$ and 
$\max(\varepsilon,\xi_{min})\le \xi_{k'}\le 0$ for $\varepsilon<0$, where 
$\xi_{min}\equiv -2t-4t'-\mu$ and $\xi_{max}=2t-4t'-\mu$ are the minimum and
the maximum over $k$ of $\xi_k$ (in the parameter range $-1<2t'/t<0$).
 
The real part of each self-energy is obtained as the Kramers-Kr\"onig 
transformation (KKT) of the corresponding imaginary part (\ref{imsigma}),
\begin{equation}
{\rm Re}\Sigma_j(k,\varepsilon)=\frac{P}{\pi}\int_{-\infty}^{+\infty}
\frac{d\omega}{\omega-\varepsilon}
{\rm Im}\Sigma_j(k,\omega){\rm sgn}(\omega).
\label{resigma}
\end{equation}
Equations (\ref{imsigma}) and (\ref{resigma}) are the starting point to obtain
single-particle physical properties such as the on-shell inverse scattering 
time, the spectral density (\ref{ak}) and the characteristic features of the
FS which will be discussed in the following sections.
  
\section{On-shell inverse scattering time.}

The inverse scattering time provides an indication of the violation of the 
normal FL behavior near criticality 
(i.e. for $M_j,T\ll t$). At the lowest order in 
perturbation theory the on-shell inverse scattering time for a quasi-particle 
with energy $\xi_k>0$ at $T=0$ is
\begin{equation}
\left( \frac{1}{\tau_k}\right)_j \equiv -2{\rm Im}\Sigma_j(k,\xi_k)
= 2\lambda_j {\nu}_j t
\int\int_{k':~0\le \xi_{k'}\le \xi_k} {d^2k'\over (2\pi /a)^2}
\frac{\xi_k-\xi_{k'}}{[\xi_k-\xi_{k'}]^2+\Omega_j^2(k-k')}
\label{tauos}
\end{equation}
for each scattering channel $j$. In this section, as a matter of illustration,
we consider the effect of a single quasi-critical mode, dropping the index $j$. 
We discuss the generic behavior of $1/\tau_k$ and individuate the regions
where the strongest violations to the FL behavior take place. When more than
one quasi-critical mode is considered, the resulting behavior is dominated
by the most singular contribution. For a given $k$ the two terms in the
denominator of (\ref{tauos}) do not vanish simultaneously unless 
$\xi_k=\xi_{k-Q}$. Then the most important contribution to (\ref{tauos}) comes 
from a region around the point $k'=k-Q$, where $\xi_{k'}=\xi_k$ and 
$\Omega(k-k')=M$. We take $\Omega(k-k')\simeq M +{\nu} a^2 
( |k-Q|^2 \phi^2 + \rho^2)/2$ where $\rho\equiv |k'|-|k-Q|\simeq 
(\xi_{k'}-\xi_{k-Q})/v_{k-Q}$, $|\rho|\ll |k-Q|$, $v_{k-Q}$ is the velocity of 
electrons at the point $k-Q$, and $\phi$ is the (small) angle between $k'$ and 
$k-Q$. The low-energy behavior of $1/\tau_k$ is most singular when
$k$ varies along the line $\xi_k=\xi_{k-Q}$, which meets the
FS at the ``hot-spot'' where $k_F\equiv k_{HS}$, such that $\xi_{k_{HS}}=
\xi_{k_{HS}-Q}=0$. Then we take $\Omega(k-k')\simeq M+{\nu} a^2 |k_{HS}-Q|^2 
\phi^2/2 $ and we perform the integral over $\rho$ in (\ref{tauos}), finding
\begin{equation}
\frac{1}{\tau_k}\simeq \frac{2\lambda t{\nu} |k_{HS}-Q|a^2\sqrt{\xi_k}}
{v_{k_{HS}-Q}}
\int_0^\infty d\theta
\log \left[ 1+\left(
 {M\over \xi_k}+ {{\nu} a^2 |k_{HS}-Q|^2 \over 2} 
\theta^2 \right)^{-2} \right],
\end{equation}  
where $\theta=\phi/\sqrt{\xi_k}$ and the upper limit is extended to $\infty$, 
to extract the leading behavior. At the QCP $M=0$ and the integration over 
$\theta$ yields the non-FL behavior \cite{hr}
\begin{equation}
\frac{1}{\tau_k} \simeq \frac{\lambda ta{\nu}}{v_{k_{HS}-Q}} 
\sqrt{\xi_k\over{\nu}}.
\label{taunfl}
\end{equation}
When $M> 0$ two different regimes exist. For $\xi_k\gg M$ the behavior 
(\ref{taunfl}) is again found, whereas for $\xi_k\ll M$ the mass term prevents 
the denominator in (\ref{tauos}) from vanishing and the FL behavior \cite{hr}
\begin{equation}
\frac{1}{\tau}
\simeq {\lambda t a \sqrt{{\nu} M}\over v_{k_{HS}-Q}}
\left( \frac{\xi_k}{M}\right)^2
\label{taufl}
\end{equation}
is recovered. When $k$ approaches $k_F$ along a line on which $\xi_k\neq 
\xi_{k-Q}$, we take $\Omega(k-k')\simeq M^*+{\nu} a^2|k_F-Q|^2\phi^2/2$,
where $M^*=M+{\nu}a^2 \xi_{k_F-Q}^2/v_{k_F-Q}^2$ is, in this case, the energy 
scale which separates the FL regime from the anomalous regime and is 
in general finite at the QCP, leading to a FL behavior for $0<\xi_K\ll M^*$. 
In this case Eq. (\ref{taufl}) holds with $M\to M^*$, $k_{HS}\to k_F$. The 
maximum violation of the FL behavior is found for $M^*=0$, i.e. $M=0$ and 
$\xi_{k_F-Q}=0$ (i.e. $k_F=k_{HS}$), 
in which case Eq. (\ref{taunfl}) holds down to $\xi_k=0$.

We study the dimensionless quantity $1/\lambda\nu\tau_k$ as a function of
$\xi_k/\xi_{max}$, according to (\ref{tauos}), in the case when $k$ 
approaches $k_{HS}$ as $\xi_k\to 0$. In this case the asymptotic behavior
(\ref{taunfl}) and (\ref{taufl}) are recovered according to the value of the
energy scale $M$ (i.e. the distance from the QCP). For the sake of definiteness 
we fix the parameters of the electronic spectrum as $t'/t=-0.25$, and the 
parameters of the fluctuation spectrum as $\nu/t=5.0$, $Q=(\pi/a,\pi/a)$. For 
this value of the characteristic wave-vector the hot spots within the first 
Brillouin zone can be determined analytically as the intersections between the 
FS ($\xi_k=0$) and the four lines $k_y=\pm\pi/a \pm k_x$. One hot spot is 
$k_{HS}=(a^{-1}{\rm arccos}\sqrt{\mu/4t'},\pi/a-a^{-1}{\rm arccos}
\sqrt{\mu/4t'})$, and the others are found by applying all the symmetry 
transformations ${\cal R}$ of the point group. We leave the ratio $\mu/t$ (or 
equivalently the hole doping $x$) as a parameter to change the position of the 
Fermi level and the location of the hot spots, and to discuss the resulting 
different regimes. We also let $M/t$ vary in the range $[10^{-6};10^{-1}]$, to 
explore the cross-over from the non-FL regime to the FL regime for $k$ 
approaching $k_{HS}$.

In Fig. \ref{fig1}(a) we take $\mu/t=-0.9$ (corresponding to 
a hole doping $x=0.17$), i.e. 
$k_{HS}=(0.32/a,2.82/a)$. As $M/t$ is reduced (by a factor of ten for each 
curve from bottom to top), the square-root behavior (\ref{taunfl})
extends over a wider range of energies. In particular, within the energy range
considered in Fig. \ref{fig1}(a), the bottom curve ($M/t=10^{-1}$) displays
a FL behavior (\ref{taufl}), whereas the top curve ($M/t=10^{-6}$) displays
a non-FL behavior (\ref{taunfl}). The intermediate curves show the crossover
from one regime to the other.

The above analysis has to be refined if the point $k_{HS}-Q$ on the FS 
is close to a singular point of the electronic spectrum so that $v_{k_{HS}-Q}$ 
is small. For instance, the characteristic wave-vector $Q_s=(\pi/a,\pi/a)$
of the AFM fluctuations connects the $M$ points $k=(\pi/a,0)$ and
$k'=(0,-\pi/a)$, or equivalently $(0,\pi/a)$, of the Brillouin zone, where 
saddle-point van Hove singularities exist at an energy $\xi_{VH}=4t'-\mu$.
A more accurate calculation is needed, as the chemical potential approaches 
${\bar\mu}\equiv 4t'$ and  $v_{k_F-Q}\sim\sqrt{|\mu-{\bar\mu}|}$, making
the estimates (\ref{taunfl}, \ref{taufl}) meaningless.
In the limiting case $\mu={\bar\mu}$, $v_{k_F-Q}=0$ and  we find 
\begin{equation}
\frac{1}{\tau_k}\simeq
8\lambda {\nu} t a^2
\int_0^{\infty}dy\int_{y\sqrt{m_x/m_y}}^{\sqrt{2m_x+m_xy^2/m_y}}dx
\frac{1-\frac{x^2}{2m_x}+\frac{y^2}{2m_y}}
{\left(1-\frac{x^2}{2m_x}+\frac{y^2}{2m_y}\right)^2+
\left[\frac{M}{\xi_k}+\frac{{\nu} a^2}{2}(x^2+y^2)\right]^2},
\label{vhlog}
\end{equation}
where $1/m_x=a^2 (2t+4t')$, $1/m_y=a^2 (2t-4t')$,
$x=k_x'/\sqrt{\xi_k}$ and $y=(k_y'-\pi/a)/\sqrt{\xi_k}$. At the QCP $M=0$ and 
(\ref{vhlog}) is independent of $\xi_k$, indicating an even stronger 
violation of the FL behavior when the hot spot coincides with a saddle
point of the electronic spectrum. 

For $M>0$ and $\xi_k\ll M$ we take $x=r\cos\theta$, $y=r\sin\theta$ and 
$z=(\frac{\cos\theta^2}{2m_x}-\frac{\sin\theta^2}{2m_y})r^2
\equiv A_{\theta}r^2$, where $A_{\theta}\simeq\eta(\theta_0-\theta)$,
 $\theta_0=\arctan\sqrt{m_y/m_x}$ and $\eta$ is a suitable constant, 
so that (\ref{vhlog}) becomes
\begin{equation}
\frac{1}{\tau_k}\simeq
4\lambda {\nu} t a^2 
\int_0^{\theta_0}\frac{d\theta}{A_{\theta}}\int_0^1dz
\frac{1-z}{(1-z)^2+
\left(\frac{{\nu} a^2}{4A_{\theta}}z+\frac{M}{\xi_k}\right)^2}
\sim\frac{\lambda a^2 t {\nu}}
{2\eta}\left(\frac{\xi_k}{M}\right)^2
\log\left(\frac{M}{\xi_k}\right).
\label{vhlog2}
\end{equation}
and the van Hove singularity introduces a logarithmic correction to the FL 
behavior. In Fig. \ref{fig1}(b) we take $\mu/t={\bar \mu}/t=-1.0$ and plot 
$1/\lambda {\nu} \tau_k $ as a function of $\xi_k/\xi_{max}$, according to 
(\ref{tauos}), for $k$ approaching $k_{HS}=(\pi/a,0)$.

As $M/t$ is reduced (by a factor of ten for each curve from bottom to top), 
the anomalous $\xi_k$-independent behavior extends over a wider range of 
energies. In particular, within the energy range considered in Fig. 
\ref{fig1}(b), the bottom curve ($M/t=10^{-1}$) displays a FL behavior 
(\ref{vhlog2}), whereas the top curve ($M/t=10^{-6}$) displays
a non-FL $\xi_k$-independent behavior. The intermediate curves show the 
crossover from one regime to the other.

The enhancement of the phase-space available for scattering makes
quasi-particles ill-defined in the vicinity of the hot spot. Moreover,
as we shall show in the following, quasi-critical scattering
near the hot spots does not only broaden the quasi-particle peak,
but also  suppresses of the polar behavior of the electron Green function
near the Fermi energy.

In the model considered in this paper, where quasi-critical charge and spin
fluctuations are simultaneously present, the hot spots corresponding to the
different modes co-exist, leading to an extension of the anomalous behavior
observed along the FS. This fact makes less stringent the objection in Ref.
\cite{hr} about the limited effect of isolated hot spots on the behavior of the
normal phase.

\section{Quasi-particle spectra} 

In this section we discuss the properties of the self-energy  
(\ref{imsigma},\ref{resigma}) at $T=0$. For the sake of clarity, we focus again
on a single quasi-critical mode, dropping the index $j$. It must, however, be 
borne in mind that, once the relevant features are individuated, the physics
of the system with coexisting charge and spin fluctuations results essentially
(though not exactly) from a superposition of the separate effects.

The imaginary part (\ref{imsigma}), at a fixed $k$, is characterized by 
different features depending on the value of the energy $\varepsilon$ compared 
to characteristic energy scales which we discuss in the following. For the
sake of the discussion, in Fig. \ref{fig2}, we fix the parameters as $t=200$
meV, $t'=-50$ meV and $\mu=-180$ meV (corresponding to a hole doping $x=0.17$)
for the electronic spectrum, $M=2\times 10^{-4}$ meV, $\nu=1000$ meV
and $Q=(\pi/a,\pi/a)$ for the fluctuation spectrum and we take a dimensionless
coupling constant $\lambda=0.2$. We also fix the external momentum
$k=(1.32/a,1.32/a)$, so that $\xi_k\simeq -10$ meV and $\xi_{k-Q}\simeq 390$ 
meV.

At energies $\varepsilon > \xi_{max}$ or $\varepsilon< \xi_{min}$ the integral 
in (\ref{imsigma}) extended over a domain which is independent of 
$\varepsilon $, so that ${\rm Im}\Sigma\simeq -C/\varepsilon$ as soon as 
$|\varepsilon|\gg \max(\xi_{max},-\xi_{min},2{\nu})$, where $C={1\over 2}
\lambda t {\nu} [1+x~{\rm sgn}(\varepsilon)]$, $x$ being the hole doping per 
unit cell. Broad dispersionless maxima for $|{\rm Im}\Sigma|$ are found at 
energies $\varepsilon\sim \min(\xi_{min},-2{\nu})$ and $\varepsilon\sim \max
(\xi_{max},2{\nu})$. Since in the case of Fig. \ref{fig2}(a) $2\nu>\xi_{max},
-\xi_{min}$, the maxima are found at energies $\varepsilon\simeq \pm 2\nu
=\pm 2000$ meV.

Within the relevant range $\xi_{min}\le \varepsilon\le \xi_{max}$, two 
characteristic energy scales exist, $M$ and $\xi_{k-Q}$. The first controls
the low-energy behavior of ${\rm Im}\Sigma(k,\varepsilon\to 0)$, which may 
be obtained by generalizing the methods discussed in the previous section, to 
the case when $\varepsilon $ varies independently of $k$. A FL behavior 
($\sim \varepsilon^2$) is found for $|\varepsilon|\ll M^*\equiv 
M+{\nu}a^2(\xi_{k-Q}/v_{k-Q})^2/2$ and a non-FL behavior
($\sim \sqrt{|\varepsilon|}$) is found for $M^*=0$ (i.e. $M=0$ and 
$\xi_{k-Q}=0$). The case shown in Fig. \ref{fig2} a corresponds to a low-energy
FL behavior .

A particular role is played by the energy $\varepsilon=\xi_{k-Q}$, since both 
terms in the denominator of (\ref{imsigma}) vanish at $k'=k-Q$ (when $M=0$). 
A peak, or a 
shoulder partially merged in the broad background structure discussed above 
is present in ${\rm Im}\Sigma$, depending on the parameters of the model.
In Fig. \ref{fig2}(a) this point is marked by a diamond at
an energy $\varepsilon\simeq 400$ meV. This dispersing structure {\sl follows}
the shadow band $\xi_{k-Q}$ with varying $k$. As $\xi_{k-Q}\rightarrow 0$ the 
peak at $\varepsilon=\xi_{k-Q}$ is suppressed, due to the fact that 
${\rm Im}\Sigma$ must vanish at the Fermi energy, but when $M=0$ the usual FL 
behavior is turned into an anomalous square-root behavior which is present all
along the curve $\xi_{k-Q}=0$, usually called shadow FS. However it must be 
pointed out that such an anomalous behavior shows up in the quasi-particle
properties only when the quasi-particle peak crosses the Fermi energy
at the shadow FS, i.e. in the vicinity of the hot spots, where 
$\xi_k\simeq\xi_{k-Q}=0$. Away from the hot spot, spectral weight is
transferred from the quasi-particle peak to the shadow peak, leading to an 
enhancement of low-laying spectral weight along the shadow FS.

The real part of the self-energy, including a correction $\delta\mu\simeq -100$ 
meV to the chemical potential, found by solving Eq. (\ref{pchim}) for
the present set of parameters, is shown in Fig. \ref{fig2} b. The presence of 
a peak in ${\rm Im}\Sigma$ at an energy $\varepsilon=\xi_{k-Q}$ produces a 
corresponding feature in ${\rm Re}\Sigma$ [marked by a diamond at 
$\varepsilon\simeq 400$ meV in Fig. \ref{fig2}(b)] which, unlike the case of
Ref. \cite{ks}, is not symmetric around $\varepsilon=\xi_{k-Q}$, due to the 
vanishing of (\ref{imsigma}) at $\varepsilon=0$. As it will become clearer in 
the discussion of the quasi-particle spectra at the end of this section,
the resulting suppression of the quasi-polar structure of ${\rm Re}\Sigma$ 
turns the shadow-bands found in Ref. \cite{ks} into incoherent resonances at 
energies $\varepsilon\simeq \xi_{k-Q}$. We also point out that, in 
correspondence of the low-energy anomalous behavior $\sim \sqrt{|\varepsilon|}$,
the suppression of the shadow feature in ${\rm Re}\Sigma$ is weaker than for 
the corresponding FL behavior ($\sim \varepsilon^2$). 

In the regime of FL violation, at low energies ${\rm Im}\Sigma=-A
{\rm sgn}(\varepsilon)\sqrt{|\varepsilon|}$ and, by KKT, ${\rm Re}\Sigma=
-A{\rm sgn}(\varepsilon)\sqrt{|\varepsilon|}+B$, where $A,B$ are
constants, which depend on the parameters of the model. The electron Green 
function looses then its polar structure. If one however insists in assigning a 
polar structure to the Green function, the wave-function renormalization factor 
$Z=\{1-[{\rm Re}\Sigma(\varepsilon)-{\rm Re}\Sigma(0)]/\varepsilon\}^{-1}
\simeq A^{-1}\sqrt{|\varepsilon|}$ vanishes at the Fermi energy. The 
quasi-particle lifetime at the hot spots, $1/\tau_{qp}= 2 Z 
|{\rm Im}\Sigma|= 2|\xi_k|$, turns out to be independent of the constants $A,B$.
This behavior is a signature of the importance of considering the appropriate
renormalizations when studying the transport properties in the presence of
singular scattering. Thus this problem turns out to be more involved than
usually considered \cite{hr}.

The single-particle spectral properties are analyzed by means of the spectral 
density (\ref{ak}). The quasi-particle peak is located, for a given $k$, at the 
intercept between the straight line $y=\varepsilon-\xi_k$ and the curve
$y={\rm Re}\Sigma(k,\varepsilon)$. This intersection, in the weak-coupling limit
$\lambda\ll 1$,  is located at an energy close to $\xi_k$
[$\varepsilon\simeq -40$ meV in Fig. \ref{fig2}(b)]. There is, however, a 
transfer of spectral weight to an energy
$\varepsilon\simeq \xi_{k-Q}$, where the denominator in (\ref{ak}) has
a local minimum [this point is marked by a diamond in Fig. \ref{fig2}(b)]. 
This incoherent spectral weight follows the dispersion of $\xi_{k-Q}$ (or
of all the $\xi_{k-Q_j}$ in the general case when more fluctuating modes are 
present) giving rise to dispersing shadow resonances, which substitute in our
approach the shadow bands found in Refs. \cite{chubukov,ks}. As an example,
in Fig. \ref{fig3} spectra are shown along the $\Gamma X$ direction in $k$ 
space, away from the hot spots, for the same set of parameters as Fig. 
\ref{fig2} and momenta $k_\ell=(\ell \pi/50 a,\ell \pi/50 a)$, with 
$\ell=15,17,19,21$. In this region of the Brillouin zone both the 
quasi-particle peak and the shadow peak have comparable intensities.
As $\xi_{k-Q}$ and $\xi_k$ get closer to each other, an increasing amount of 
spectral weight is transferred from the quasi-particle peak to the shadow peak.

\section{Spectral density and Fermi surface in ARPES experiments.}

In this section we use the general results obtained above to discuss recent 
ARPES experiments Bi2212 near optimal doping 
\cite{bianconi,marshall,larosa,arpes,saini}. The increasing experimental
evidence for a stripe phase \cite{xxx,ics}, reinforces our idea of having a 
charge modulation which enslaves a spin modulation, leading to the model
(\ref{hamilt}) with coexisting charge and spin fluctuations. Thus, we
deal here with the full self-energy (\ref{selfene}) associated with critical
modes of both types. We point out again that, when more than one mode is 
considered, the resulting effects are essentially a superposition of those
discussed in the previous sections for a single generic mode. The presence of 
AFM fluctuations triggered by the incipient stripe instability is described,
within our model, by the condition that both modes are quasi-critical, i.e.
$M_c\le M_s\ll t$. We neglect in the following the possibility for a modulation
of the AFM characteristic wave-vector induced by stripe formation and the
related splitting \cite{tranq}, which adds minor changes to the quantities 
under discussion, and does not appear to be resolved by ARPES data.

We make a comparison between the spectral density (\ref{ak}) and the 
experimental energy distribution curves (EDCs) \cite{notasaini}
and we analyze the shape and 
properties of the FS. To reproduce typical conditions in ARPES experiments we 
introduce a convoluted spectral density
\begin{equation}
\tilde{A}_R(k,\varepsilon)=\int_{-\infty}^{+\infty}d\varepsilon'
A(k,\varepsilon')f(\varepsilon') {\cal E}_R (\varepsilon'-\varepsilon)
\label{convak}
\end{equation}
which takes care of the absence of occupied states above the Fermi energy,
through the Fermi function, and of the experimental energy resolution $R$, 
through the resolution function ${\cal E}_R(\varepsilon)=
\exp(-\varepsilon^2/2R^2)/\sqrt{2\pi R^2}$. As it is customarily in ARPES 
experiments, the resolution in momentum space is assumed to be finer than the 
energy resolution, so that no convolution over $k$ is performed in 
(\ref{convak}). A comparison is then possible between
the experimental EDCs and the convoluted density (\ref{convak}) as 
a function of $\varepsilon$ at given $k$. The self-energy which enters the
expression (\ref{ak}) is evaluated here with a smooth cut-off $\Delta=0.4$, 
which selects transferred momenta close to the characteristic wave-vectors
$Q_c,Q_s$, to improve the agreement with experimental results [see Eq. 
(\ref{imsigma}) and the subsequent discussion].

A particular care is required to discuss the experimentally determined FS. 
Indeed, the presence of the shadow peaks and of the associated incoherent 
spectral weight at low energies along the shadow FS, makes the quasi-particle 
FS [determined by the equation $\xi_k+{\rm Re}\Sigma(k,\varepsilon=0)=0$]
not appropriate. The ARPES EDCs reveal the FS via the zero-energy crossing of 
the dispersing intensity peaks, regardless of their coherence. More recently 
the FS was accurately
determined as the distribution of spectral weight within an energy 
window around the Fermi energy, by means of the angle scanning photoemission 
(ASP) \cite{bianconi}. Once again the experimental distinction between the main 
and the shadow FS is lost. Thus, we associate with each $k$ point 
the integrated spectral weight of low-laying occupied states 

\begin{equation}
p_k^W=\int_{-W}^W A(k,\varepsilon)f(\varepsilon)
d\varepsilon,
\label{pesi}
\end{equation}
where $W\ll t$ \cite{notap}, and we compare the distribution of $p_k^W$ with 
the FS observed in ASP experiments \cite{notafs}. We point out that, in the 
absence of an anomalous transfer of spectral weight to low energies, such a 
determination of the FS coincides essentially with the theoretical definition 
in terms of quasi-particle, at least for reasonably small W.

We want to discuss the evolution of the band structure as determined by 
following the intensity peaks in ARPES experiments and the properties of the
FS, which is determined with more details in ASP experiments. 
To reproduce the experimental conditions in ARPES experiments \cite{marshall}
we take $T=10$ meV and $R=15$ meV in (\ref{convak}). To reproduce the 
conditions in ASP experiments \cite{bianconi} we take $T=25$ meV and $W=25$
meV in (\ref{pesi}).
 
The parameters of the free-electron band $\xi_k$ will be fixed to reproduce the 
band structure, the FS and the electron filling of Bi2212 at optimal doping, 
i.e. $t= 200$ , $t'=-50$ meV, $\mu=-180$ meV, corresponding to a hole doping 
$x=0.17$.

The experimentally observed FS in Bi2212 is approximately mirror-symmetric with 
respect to the $\Gamma X(Y)$ axes, but not with respect to the 
$\Gamma M(M_1)$ axes \cite{bianconi}. An additional feature of the experiments
is the presence of a dispersing band along the $\Gamma M$ direction, crossing
the Fermi energy at $\tilde k_F=(0,0.2\pi/a)$, which is absent along the 
$\Gamma M_1$ direction \cite{saini}.

Within the stripe-QCP scenario, the symmetry of the FS is expected to be 
lowered by the interaction between electrons and charge fluctuations, 
characterized by an incommensurate wave-vector $Q_c$, whereas in the presence
of spin fluctuations only, the excitation spectrum and the FS are expected
to have the full symmetry of the lattice, due to the peculiar commensurability 
of the characteristic wave-vector of the N\'{e}el AFM structure, 
$Q_s=(\pi/a,\pi/a)$. Within our model, in order to preserve the symmetry 
$k\to -k$, we always perform an average over $\pm Q_c$, i.e. we take a 
self-energy ${\bar \Sigma}_c={1\over 2}[\Sigma_c(Q_c)+\Sigma_c(-Q_c)]$.
Other point symmetries of the original system may be broken according to the
direction of $Q_c$, which is system \cite{tranq,mook} and model dependent 
\cite{goetz}. 

Based on the symmetry of the FS observed in Bi2212 \cite{bianconi}, for this
specific system we take $Q_c\simeq(0.4\pi/a,-0.4\pi/a)$, directed along the 
$\Gamma Y$ direction.

We discuss in some detail the case $\nu_c=\nu_s=200$ meV, $M_c=M_s=10$ meV and
$\lambda_c=1/2$, $\lambda_s=1/6$ (to which Fig. \ref{fig4} refers). 
The correction $\delta\mu$ to the chemical potential, obtained by solving
Eq. (\ref{pchim}) for the present set of parameters, is $\delta\mu=-40$ meV.

The single-particle spectra are characterized by the presence of shadow peaks
associated with $\pm Q_c$ and $Q_s$. The two shadow peaks, associated with
$\xi_{k\pm Q_c}$, and the corresponding branches of the shadow FS 
($\xi_{k\pm Q_c}=0$) do not coincide, as it is instead the case for the
highly commensurate vector $Q_s$ (which is equivalent to $-Q_s$). 

In Fig. \ref{fig4} the EDCs along the $\Gamma M$ direction are reported in the
panel on the right. The momentum $k$ is increased from top to bottom by a
uniform step $\Delta k=\pi/8 a$. The quasi-particle peak moves from the left 
to the right towards the Fermi energy and interferes with the shadow peak 
associated with $+Q_c$ (see the third curve from the top) which moves 
initially to lower energies (fourth curve) and then again towards the Fermi 
energy. This shadow peak appears in Fig. \ref{fig4} as a shoulder merged in the 
quasi-particle peak in the  seventh curve from the top.

The shadow peak associated with $Q_s$ is located at high energies above the 
Fermi level, near the $\Gamma$ point. However, as $k$ is increased, this peak 
moves towards the Fermi energy, while the quasi-particle looses spectral
weight, which is transferred to the shadow peaks. In the last three curves
in Fig. \ref{fig4} the shadow peak associated with $Q_s$ appears as a broad
incoherent peak moving towards an energy $\varepsilon\simeq -150$ meV while
approaching the $M$ point. We associate this peak with a corresponding
feature observed in Bi2212 at an energy $\varepsilon \simeq -200$ meV 
\cite{marshall}. 

For a comparison we report, in Figs. \ref{fig5} and \ref{fig6}
respectively, the effects due to spin and charge modes when considered 
separately. In the right panel of Fig. \ref{fig5}, from top to bottom, the
quasi-particle peak looses its weight as it moves towards the Fermi energy, 
while near the $M$ point a broader shadow structure still appears below the 
Fermi energy. Recently this feature was discussed within the NAFL scenario in 
Ref. \cite{schpi}, where a full re-summation of the diagrammatic expansion in 
the static limit was performed. We point out, however, that these results match 
those obtained within our simpler perturbative approach. In the right panel
of Fig. \ref{fig6} we report the single-particle spectra in the $\Gamma M$
direction in the presence of charge fluctuations only. The first six curves 
from the top are very similar to the corresponding curves in Fig. \ref{fig4}, 
since there the spin-related shadow peak is located at much higher energy 
above the Fermi level. The shadow peak related to $+Q_c$ appears to cross the 
Fermi level at ${\tilde k}_F\simeq (2.4/a,0)$ in Fig. \ref{fig6}. In the last
three curves, due to the absence of the effects related to spin fluctuations, 
the quasi-particle peak never reaches the Fermi energy along the $\gamma M$ 
direction. The van Hove singularity is located at $\xi_{VH}\simeq -50$ meV. 

The description of the additional feature observed only in the $\Gamma M$
direction \cite{saini} would require the introduction of an additional charge
modulation wit a wave-vector $Q_c'\simeq (0,2\pi/3a)$. We do not discuss these
finer details here.

We now discuss the properties of the FS according to the distribution 
(\ref{pesi}) of low-laying spectral weigh. The main features of the FS,
reported in the left panel of Fig. \ref{fig4}, are an inhomogeneous
distribution of spectral weight, which is maximum near the diagonals of the
Brillouin zone, and an asymmetric suppression of spectral weight near the
$M(M_1)$ points, as experimentally observed \cite{bianconi}. The first
aspect is related to the evolution of the FS in the underdoped regime, as
the temperature is lowered down to the superconducting critical temperature
\cite{norman}. We point out that, since we are at a fixed doping $x=0.17$, and 
since our model (\ref{hamilt}) is inadequate to discuss the underdoped regime,
where the spectrum of charge fluctuations is presumably modified by the
presence of preformed Cooper pairs, we can only describe the onset of this
evolution. The second aspect is understood as a co-operative effect of spin 
fluctuations, which lead to a symmetric suppression (see the left panel in
Fig. \ref{fig5}, where spin fluctuations only are considered), and charge
fluctuations, which introduce an asymmetric modulation related to the direction
and magnitude of $Q_c$ (see the left panel in Fig. \ref{fig6}, where charge 
fluctuations only are considered).

The branches of the shadow FS due to charge fluctuations, which are clearly
visible in Fig. \ref{fig6}, where they cross transversally the $\Gamma M(M_1)$
directions in the vicinity of the $M(M_1)$ points, interfere with the branches 
due to spin fluctuations, when both mode are considered as in Fig. \ref{fig4}.

Another feature of the FS, related to spin fluctuations, is the possibility for
the presence of hole pockets around the points $(\pm \pi/2a,\pm \pi/2a)$ of the 
Brillouin zone. According to our interpretation of the experimentally observed
FS we choose a set of parameters which leads to very weak hole pockets, which
are well below the threshold fixed for the representative points in Figs. 
\ref{fig4} and \ref{fig5}, and are not reported. Larger $|t'/t|$, and/or
larger ${\nu}_s/t$, and/or a sharper cut-off $\Delta$, and/or a smaller energy 
window $W$, make this phenomenon more pronounced> Whenever present, the hole 
pockets are not associated with a topological change of the quasi-particle
FS, as in Ref. \cite{chubukov}, but rather correspond to a distribution
$p_k^W$ which is characterized by local maxima at the FS ($\xi_k=0$)
and at the branch of the shadow FS ($\xi_{k-Q_s}=0$), and minima at 
$(\pm \pi/2a,\pm \pi/2a)$. The experimental situation
is still controversial within this respect. 
Shadow bands shifted by a wave-vector $Q_s$ and/or the related hole pockets
were reported in Refs. \cite{bianconi,larosa,aebi} but were not detected in 
Refs. \cite{marshall,norman}.

\section{Conclusions}

We studied the changes in the single-particle properties due to
the coupling of electrons to quasi-critical charge and spin fluctuations in 
the quantum critical region around a stripe QCP. Violations to the normal FL 
behavior appear in the on-shell inverse scattering time, at those points of the
FS connected by a characteristic wave-vector of the critical fluctuations
(hot spots). The violations are stronger when a hot spot is located near a 
singular point of the electronic spectrum, leading to a finite inverse 
scattering time. This result may be relevant for the physics of high-$T_c$ 
cuprates, where extended van Hove singularities have been repeatedly observed 
\cite{vhs} close to the Fermi energy at the $M$ points, in the presence of 
critical fluctuations characterized by either a large ($\sim Q_s$) or a small 
$Q_c$ wave-vector, leading to hot spots in the vicinity of the $M$ points.

The study of the quasi-particle spectra showed that the violation 
to the FL behavior are associated with the transfer of spectral weight from the 
quasi-particle band to incoherent shadow peaks. These feature are
dispersing and show up in the ARPES EDCs, though they do not correspond to
poles of the electron Green function. As the shadow peaks approach the
Fermi level, they produce an enhancement of low-laying spectral weight
which is seen as a shadow FS within a finite energy resolution.

The evolution of the FS is thus associated with the change in the 
distribution of the low-laying spectral weight and not with the
topological modification of the quasi-particle FS proposed in Ref.
\cite{chubukov}.

The quasi-particle spectra and the FS obtained within the model with 
coexisting charge and spin fluctuations capture the relevant features observed 
in ARPES experiments on optimally doped Bi2212, and namely the presence of 
broad features in the EDCs (besides the quasi-particle peak) the inhomogeneous 
distribution of spectral weigh along the FS and the asymmetric suppression of 
spectral weight around the $M(M_1)$ points of the Brillouin zone 
\cite{notaumk}. 

On the other hand we suggest that the ARPES experiments may be used to extract 
informations about the parameters entering in the fluctuation spectra 
(\ref{chi}), which can then be tested consistently within the theory. In 
particular, the direction and magnitude of $Q_c$ might be deduced from the 
modulation in the suppression of spectral weight along branches of the FS, as 
suggested in Ref. \cite{bianconi}, or from a systematic, and still lacking,
investigation of the related shadow peaks.

{\bf Acknowledgements:} Part of this work was carried out with the financial 
support of the INFM, PRA 1996. The authors would like to thank
Prof. C. Castellani for many useful discussions and suggestions.

\newpage

{\bf {\centerline {FIGURE CAPTIONS}}}

\begin{figure}[htbp]   
    \begin{center}
       \setlength{\unitlength}{1truecm}
       \begin{picture}(5.0,15.0)
          \put(-6.0,-8.0){\epsfbox{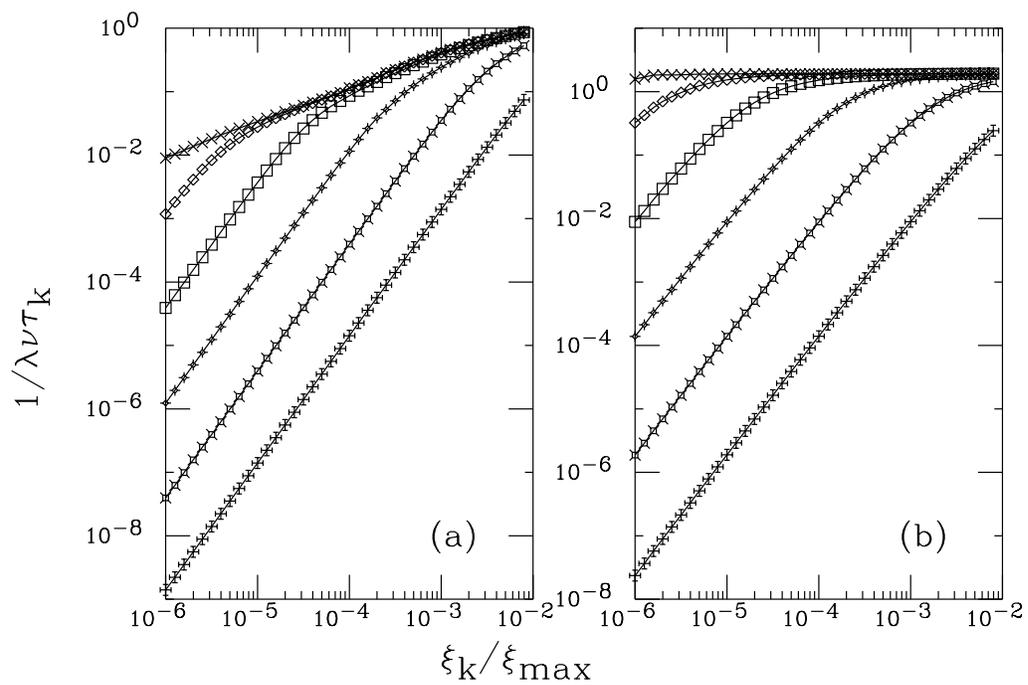}}
       \end{picture}
    \end{center}
    \caption{S. Caprara {\sl et al.} - {\it  Single-particle properties of a model ...}}
    \protect\label{fig1}
% ...
\end{figure}

Fig. \ref{fig1}: Dimensionless on-shell inverse scattering time
at the hot spot, as a function of $\xi_k/\xi_{max}$. 
(a) The parameters of the electronic
spectrum are $t'/t=-0.25$, $\mu/t=-0.9$ (corresponding to a hole
doping $x=0.17$); the parameters of the fluctuation spectrum
are $\nu/t=5.0$, $Q=(\pi/a,\pi/a)$, while $M/t$ varies in the range
$[10^{-6};10^{-1}]$ increasing by a factor of 10 for each curve from
top to bottom. The hot spot is located at $k_{HS}=(0.32/a,2.82/a)$.
(b) Same parameters as in (a), except for the chemical
potential $\mu/t=-1.0$ (corresponding to a hole doping $x=0.22$). The
van Hove singularity in the electronic spectrum is located at the Fermi level.
The hot spot is located at $k_{HS}=(\pi/a,0).$
\par
\begin{figure}[htbp]   
    \begin{center}
       \setlength{\unitlength}{1truecm}
       \begin{picture}(5.0,15.0)
          \put(-6.0,-8.0){\epsfbox{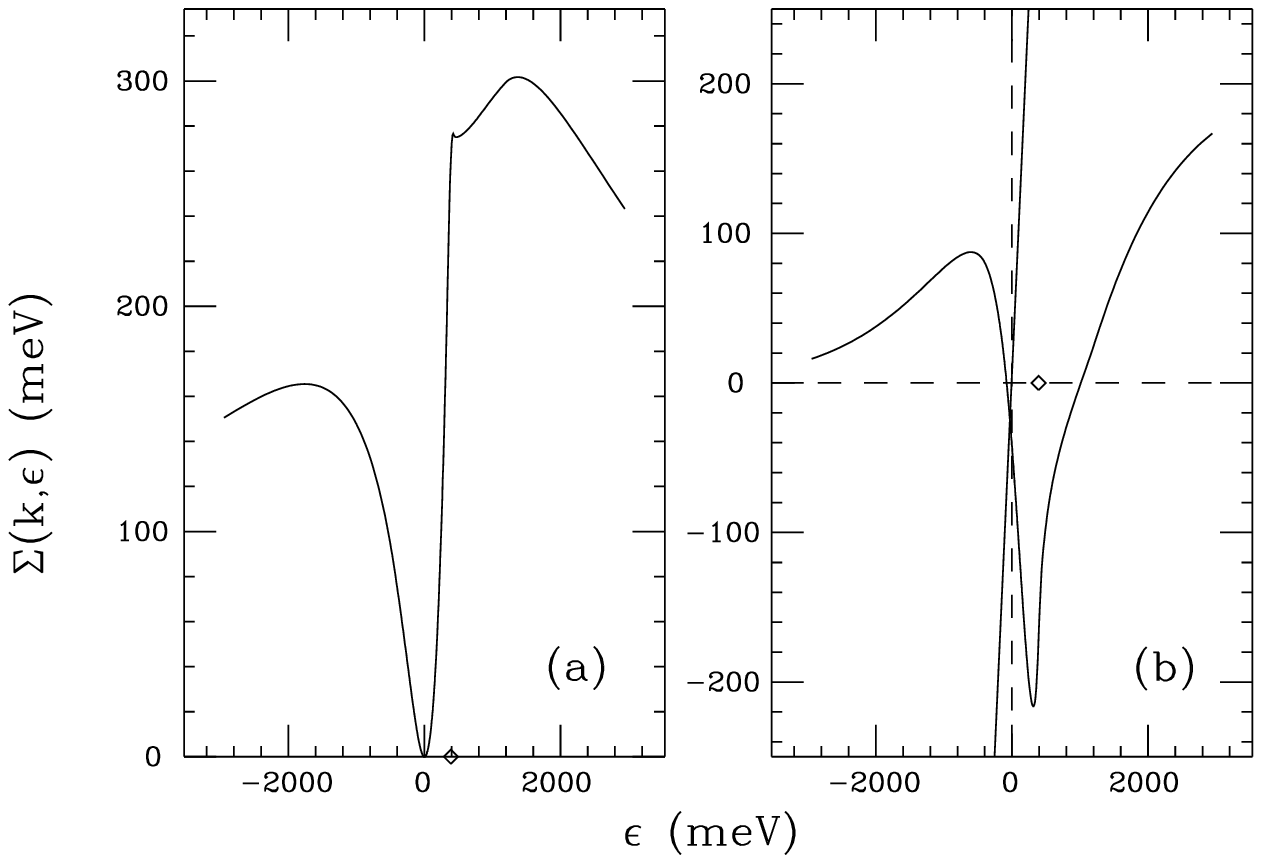}}
       \end{picture}
    \end{center}
    \caption{S. Caprara {\sl et al.} - {\it  Single-particle properties of a model ...}}
    \protect\label{fig2}
% ...
\end{figure}

Fig. \ref{fig2}: 
Self-energy as a function of the external energy $\varepsilon$. The
parameters are the same as in Fig. \ref{fig1}(a), with $t=200$ meV,
to fix the energy scale, and $M/t=10^{-6}$.
The dimensionless coupling is $\lambda=0.2$ and
the momentum is $k=(1.32/a,1.32/a)$. For these values of the parameters
$\xi_k\simeq-10$ meV and $\xi_{k-Q}\simeq 390$ meV (marked by a diamond
on the energy axis). (a) Imaginary part. The broad dispersionless 
maxima are located at $\varepsilon\simeq\pm 2000$ meV. The shadow
feature is located at $\varepsilon=\xi_{k-Q}$. (b) Real part, including
a correction to the chemical potential $\delta\mu\simeq -100$ meV. 
The shadow feature is located at $\varepsilon=\xi_{k-Q}$. 
The quasi-particle
peak corresponds to the intercept between the solid line $\varepsilon-\xi_k$
and the curve ${\rm Re}\Sigma(k,\varepsilon)$, and is located
at an energy $\varepsilon\simeq -50$ meV. The dashed lines represent the axes.
\par
\begin{figure}[htbp]   
    \begin{center}
       \setlength{\unitlength}{1truecm}
       \begin{picture}(5.0,15.0)
          \put(-6.0,-8.0){\epsfbox{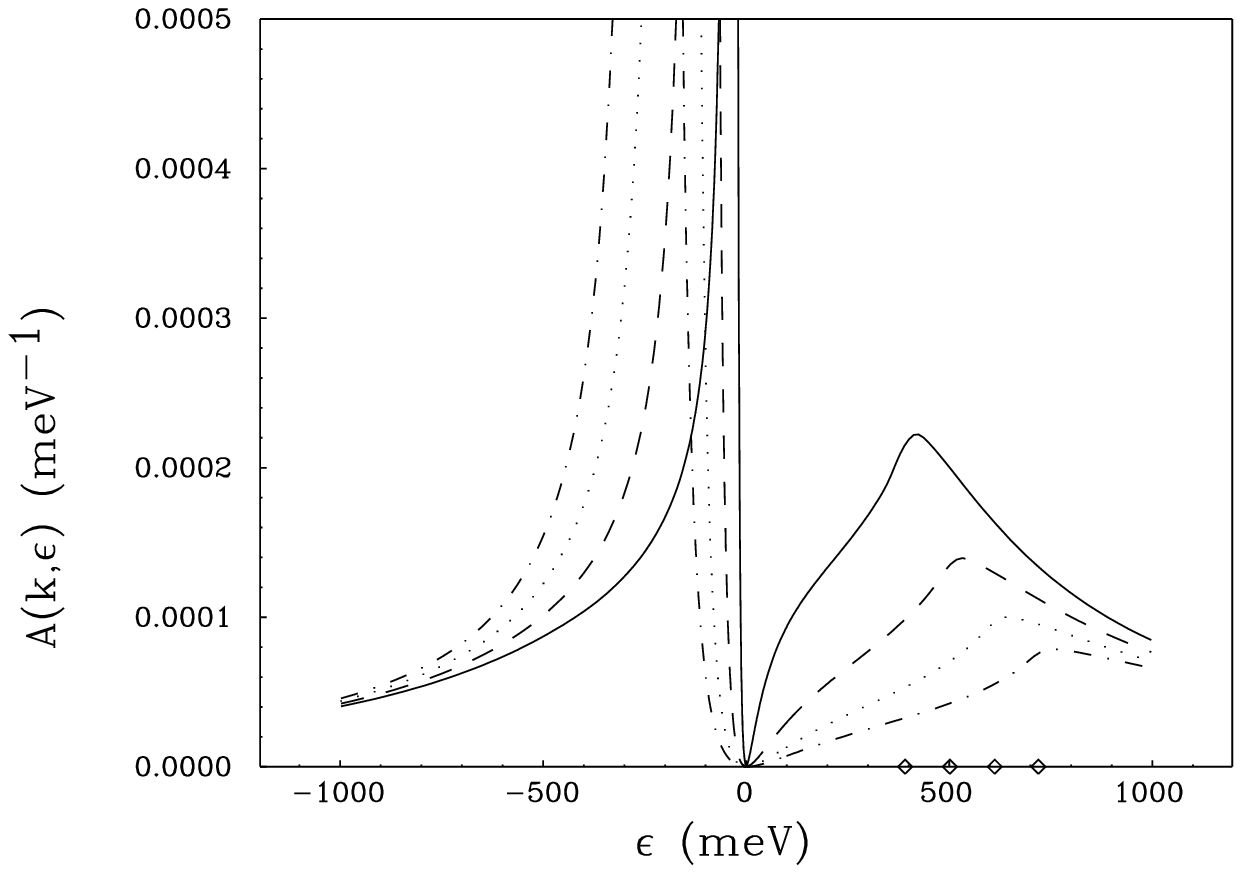}}
       \end{picture}
    \end{center}
    \caption{S. Caprara {\sl et al.} - {\it  Single-particle properties of a model ...}}
    \protect\label{fig3}
% ...
\end{figure}

Fig. \ref{fig3}: 
Spectral density for $k_\ell=(\ell\pi/50a,\ell\pi/50a)$, 
with $\ell=15,17,19,21$ (dot-dashed, dotted, dashed and solid lines
respectively). The other parameters are the same as in Fig. \ref{fig2}. 
The quasi-particle
peak is moving from the left to the right (towards the Fermi energy)
as $\ell$ is increased. The shadow peak is moving from the right to the
left as $\ell$ is increased. The position of the shadow band, 
$\varepsilon=\xi_{k-Q}$,
is marked by diamonds on the energy axis. Spectral weight is transferred
from the quasi-particle peak (on the left) to the shadow peak 
(on the right) as the two peak get closer to each other.
\par
\begin{figure}[htbp]   
    \begin{center}
       \setlength{\unitlength}{1truecm}
       \begin{picture}(5.0,15.0)
          \put(-6.0,-8.0){\epsfbox{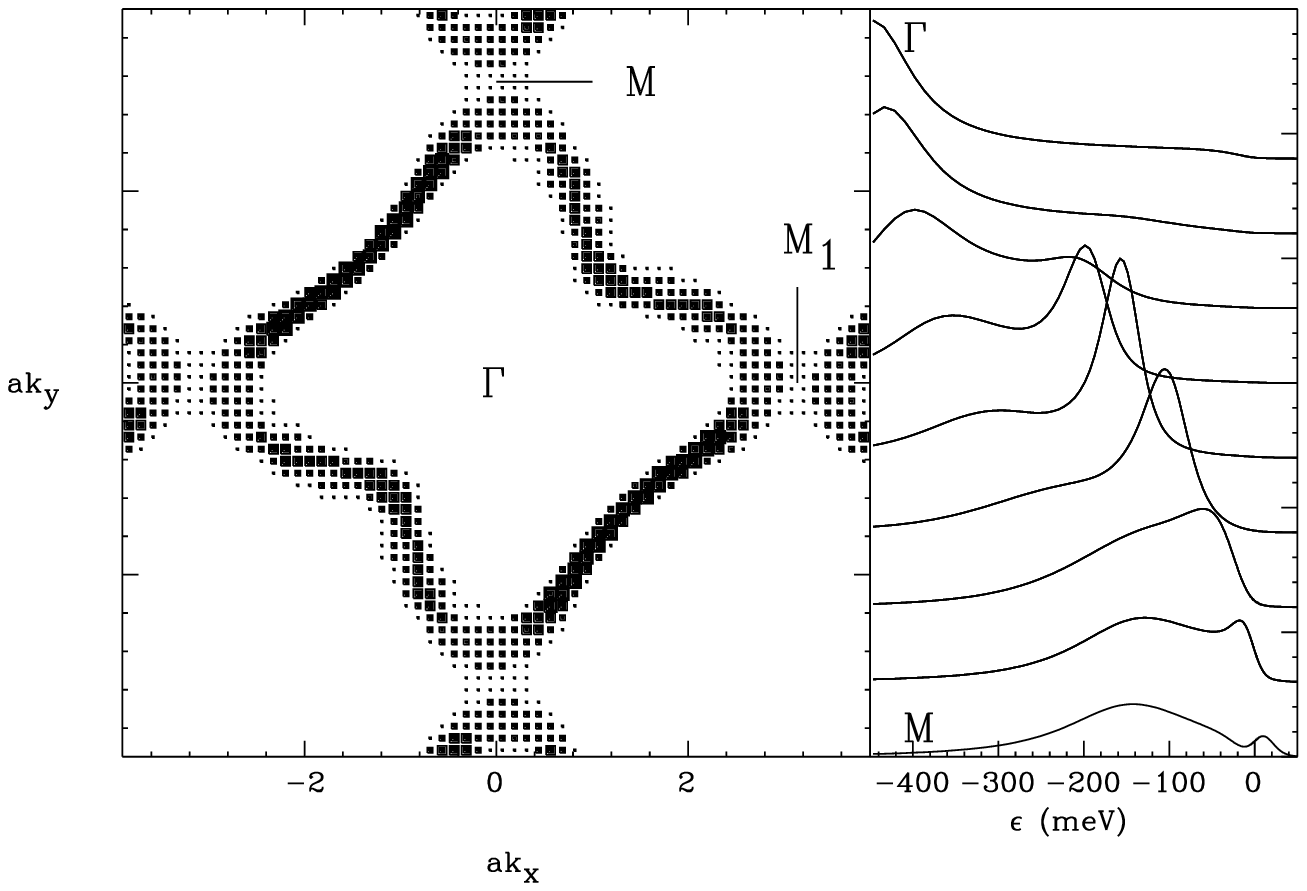}}
       \end{picture}
    \end{center}
    \caption{S. Caprara {\sl et al.} - {\it  Single-particle properties of a model ...}}
    \protect\label{fig4}
% ...
\end{figure}

Fig. \ref{fig4}: 
LEFT: Distribution of low-laying spectral
weight $p_k^W$ within the Brillouin zone in the case of electrons
coupled to both charge and spin fluctuations. 
The weight is reduced by a factor of 2 as the size of the black squares is 
reduced. Four classes are shown. The weight less than $1/16$ of the maximum
is not reported.
The parameters of the electronic spectrum
are $t=200$ meV, $t'=50$ meV, $\mu=-180$ meV (corresponding to a hole doping 
$x=0.17$). The parameters of the
fluctuation spectra are $\nu_{s,c}=200$ meV, $M_{s,c}=10$ meV, $Q_s=(\pi/a,
\pi/a)$ and $Q_c=(0.4\pi/a,-0.4\pi/a)$. 
The dimensionless coupling constants are $\lambda_s=1/6$,
$\lambda_c=1/2$ and the
dimensionless cut-off is $\Delta=0.4$. The correction to the chemical
potential is $\delta\mu\simeq-40$ meV. The integrated spectral weight
$p_k^W$ is calculated according to (\ref{pesi}) with $W=25$ meV
and $T=25$ meV. RIGHT: Convoluted spectral density $\tilde A_R(k,\varepsilon)$,
calculated according to (\ref{convak}) with $R=10$ meV and $T=10$ meV,
along the $\Gamma M$ direction for the same parameters as in the
left panel. The momentum is uniformly increased from top to bottom.
\par
\begin{figure}[htbp]   
    \begin{center}
       \setlength{\unitlength}{1truecm}
       \begin{picture}(5.0,15.0)
          \put(-6.0,-8.0){\epsfbox{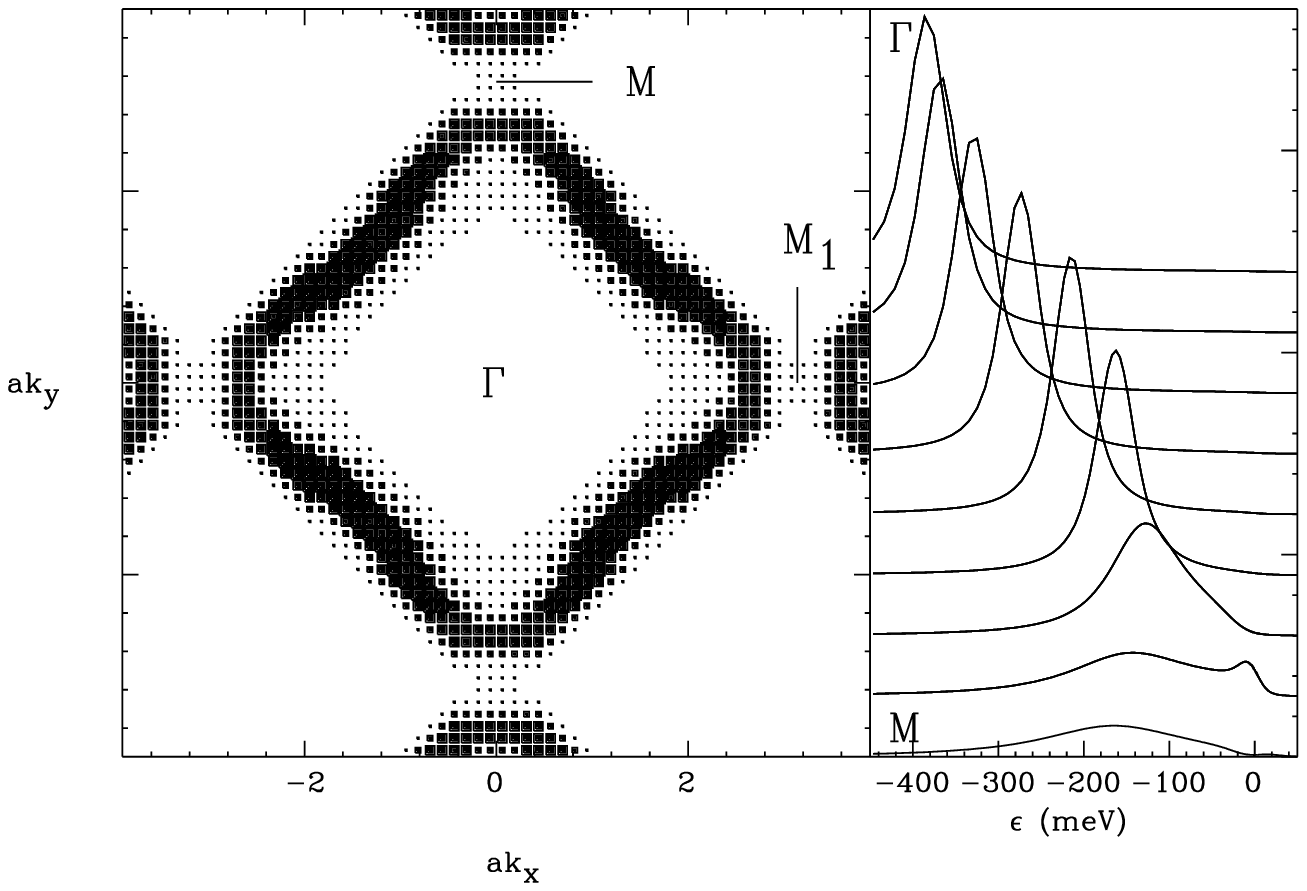}}
       \end{picture}
    \end{center}
    \caption{S. Caprara {\sl et al.} - {\it  Single-particle properties of a model ...}}
    \protect\label{fig5}
% ...
\end{figure}

Fig. \ref{fig5}: 
LEFT: Distribution of low-laying spectral
weight $p_k^W$ within the Brillouin zone in the case of electrons
coupled to spin fluctuations only. 
The weight is reduced by a factor of 2 as the size of the black squares is 
reduced. Four classes are shown. The weight less than $1/16$ of the maximum
is not reported.
The parameters of the electronic spectrum
are the same as in Fig. \ref{fig4}. The parameters of the
spin-fluctuation spectrum are $\nu_s=200$ meV, $M_s=10$ meV, $Q_s=(\pi/a,
\pi/a)$. The dimensionless coupling constant is $\lambda_s=1/3$ and the
dimensionless cut-off is $\Delta=0.4$. The correction to the chemical
potential is $\delta\mu=-70$ meV. The integrated spectral weight
$p_k^W$ is calculated according to (\ref{pesi}) with $W=25$ meV
and $T=25$ meV. RIGHT: Convoluted spectral density $\tilde A_R(k,\varepsilon)$,
calculated according to (\ref{convak}) with $R=10$ meV and $T=10$ meV,
along the $\Gamma M$ direction for the same parameters as in the
left panel. The momentum is uniformly increased from top to bottom.
\par
\begin{figure}[htbp]   
    \begin{center}
       \setlength{\unitlength}{1truecm}
       \begin{picture}(5.0,15.0)
          \put(-6.0,-8.0){\epsfbox{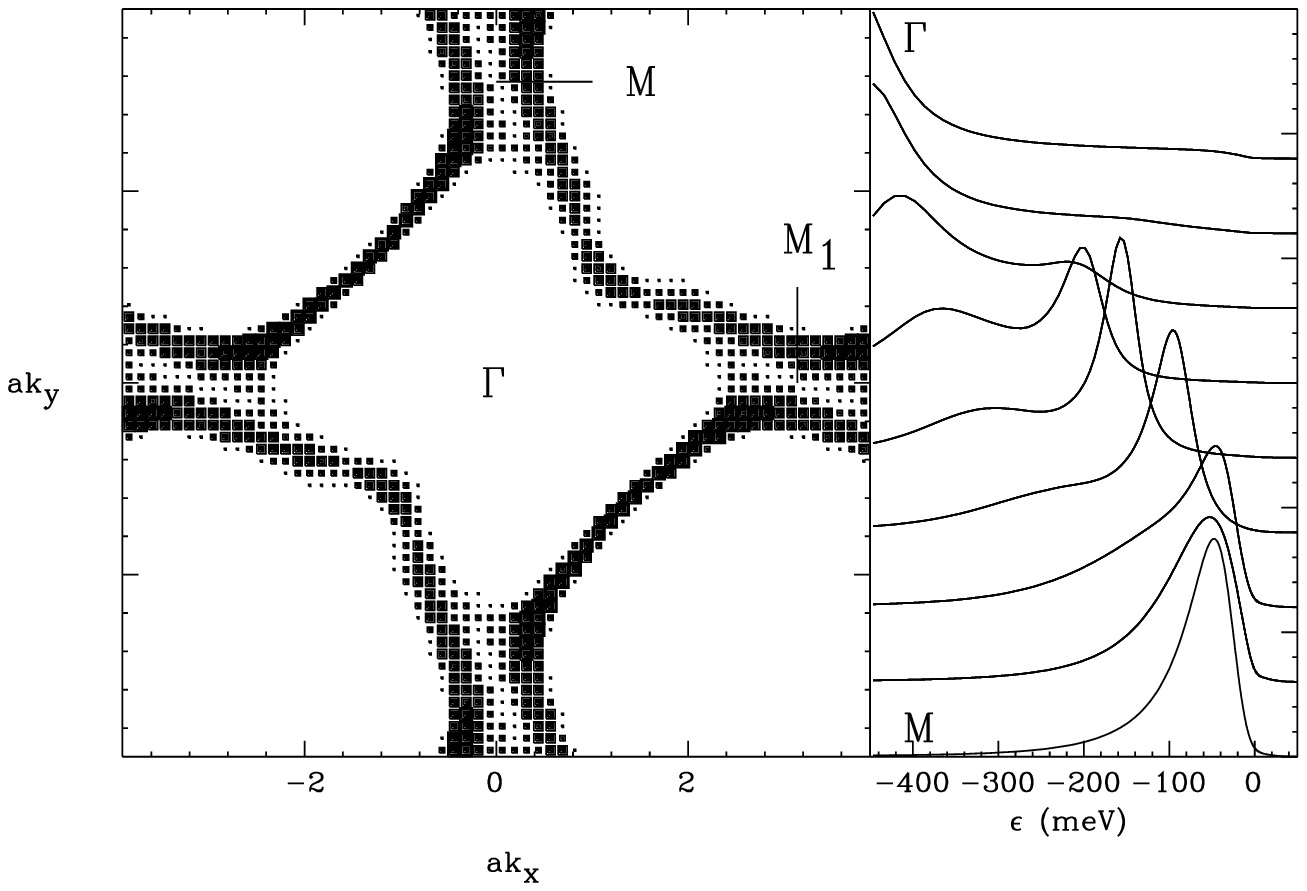}}
       \end{picture}
    \end{center}
    \caption{S. Caprara {\sl et al.} - {\it  Single-particle properties of a model ...}}
    \protect\label{fig6}
% ...
\end{figure}

Fig. \ref{fig6}: 
LEFT: Distribution of low-laying spectral
weight $p_k^W$ within the Brillouin zone in the case of electrons
coupled to charge fluctuations only. 
The weight is reduced by a factor of 2 as the size of the black squares is 
reduced. Four classes are shown. The weight less than $1/16$ of the maximum
is not reported.
The parameters of the electronic spectrum
are the same as in Fig. \ref{fig4}. The parameters of the
charge-fluctuation spectrum are $\nu_c=200$ meV, $M_c=10$ meV, $Q_c=(0.4\pi/a,
-0.4\pi/a)$. The dimensionless coupling constant is $\lambda_c=1/2$ and the
dimensionless cut-off is $\Delta=0.4$. The correction to the chemical
potential is negligible. The integrated spectral weight
$p_k^W$ is calculated according to (\ref{pesi}) with $W=25$ meV
and $T=25$ meV. RIGHT: Convoluted spectral density $\tilde A_R(k,\varepsilon)$,
calculated according to (\ref{convak}) with $R=10$ meV and $T=10$ meV,
along the $\Gamma M$ direction for the same parameters as in the
left panel. The momentum is uniformly increased from top to bottom.


\begin{thebibliography}{99}
\bibitem{notaks} We point out that the
theory proposed in Ref. \cite{ks} is based on a modification of the
mean-field susceptibility, and not of the mean-field self-energy. However
this difference is unessential as far as a qualitative discussion of the
properties of the resulting quasi-particle spectra is concerned.
\bibitem{notap}
We assume here a different meaning for the experimental resolution
in ARPES and ASP experiment. It is, however, evident that
$p^W_k\simeq \sqrt{2\pi W^2} \tilde A_{R=W}(k,\varepsilon=0)$.
\bibitem{notafs}
In more detail,
such weights are grouped into classes according to the value of the class index
$i=[1-\log _b (p^W_k/p^W_{max})]$,
where the square brackets select the integer part, $b$ is a suitably
chosen basis and $p^W_{max}$ is the maximum over $k$ of
$p^W_k$. Each value of $i=1,2,....$ corresponds to a weight
$p^W_{max}/b^i<p^W_k\le p^W_{max}/b^{i-1}$. We plot points with
$i\le i_0$, where $i_0$ represents the threshold
of the experimental apparatus.
\bibitem{notaumk}
We are aware of the fact that the intrinsic modulation
of the Bi-planes in the $(1,-1)$ direction introduces satellites
branches of the FS \cite{umkl} due to the umklapp of photo-electrons which are
scattered by the periodic Bi-planes before reaching the detector. 
However such an effect seems not to be sufficient to explain the asymmetric 
distribution of spectral weight around the $M(M_1)$ points, since
the weight of the satellites branches is too small to produce the observed 
intensities.
\end{thebibliography}
\end{document}